\documentstyle[emulateapj]{article}

\def\simlt{\lower.5ex\hbox{$\; \buildrel < \over \sim \;$}}
\def\simgt{\lower.5ex\hbox{$\; \buildrel > \over \sim \;$}}
\def\gsim{\;\rlap{\lower 2.5pt
\hbox{$\sim$}}\raise 1.5pt\hbox{$>$}\;}
\def\lsim{\;\rlap{\lower 2.5pt
   \hbox{$\sim$}}\raise 1.5pt\hbox{$<$}\;}

\def\mpy{ \ M_\odot \, \rm yr^{-1}}
\def\spose#1{\hbox to 0pt{#1\hss}}
\def\lta{\mathrel{\spose{\lower 3pt\hbox{$\mathchar''218$}}
     \raise 2.0pt\hbox{$\mathchar''13C$}}}
\def\gta{\mathrel{\spose{\lower 3pt\hbox{$\mathchar''218$}}
     \raise 2.0pt\hbox{$\mathchar''13E$}}}
\newcommand{\beq}{\begin{equation}}
\newcommand{\eeq}{\end{equation}}


\lefthead{MENOU \& QUATAERT}
\righthead{{INSTABILITIES IN SUPERMASSIVE BLACK HOLE DISKS}}

\begin{document}

\title{Ionization, Magneto-rotational and Gravitational Instabilities in Thin Accretion Disks Around Supermassive Black Holes}

\author{Kristen Menou\altaffilmark{a,1} \& Eliot Quataert\altaffilmark{b,1}}

\affil{$^{a}$Princeton University, Department of Astrophysical Sciences, Princeton, NJ 08544; kristen@astro.princeton.edu}
\affil{$^{b}$Institute for Advanced Study, School of Natural Sciences, Einstein Drive, Princeton, NJ 08540; eliot@ias.edu}

\altaffiltext{1}{Chandra Fellow}

\authoremail{kmenou@astro.princeton.edu}
\authoremail{eliot@ias.edu}

\vspace{\baselineskip}

\begin{abstract}
We consider the combined role of the thermal ionization,
magneto-rotational and gravitational instabilities in thin accretion
disks around supermassive black holes.  We find that in the portions
of the disk unstable to the ionization instability, the gas remains
well coupled to the magnetic field even on the cold, neutral branch of
the thermal limit cycle.  This suggests that the ionization
instability is not a significant source of large amplitude
time-dependent accretion in AGN.  We also argue that, for $\dot M
\gsim 10^{-2} \mpy$, the gravitationally unstable and
magneto-rotationally unstable regions of the accretion disk overlap;
for lower accretion rates they may not.  Some low-luminosity AGN,
e.g. NGC~4258, may thus be in a transient phase in which mass is
building up in a non-accreting gravitationally and
magneto-rotationally stable ``dead zone.''  We comment on possible
implications of these findings.

\end{abstract}

{\it subject headings}: accretion, accretion disks -- black hole
physics -- MHD -- turbulence -- galaxies: nuclei -- quasars: general

\section{Introduction}

Thin accretion disks around compact objects are locally unstable at
central temperatures $\sim 10^4$ K, due to large opacity changes when
hydrogen is partially ionized (Meyer \& Meyer-Hofmeister 1981).  It is
generally believed that this instability drives the outbursts of dwarf
novae (DN; Smak 1984; Cannizzo 1993 and references therein) and soft
X-ray transients (SXTs; van Paradijs \& Verbunt 1984; Mineshige \&
Wheeler 1989; Dubus et al. 2000).  Physically, the disk cannot accrete
at a steady rate because a range of radii are thermally and viscously
unstable.  As a result the disk alternates between two stable branches
(see \S2): (1) a hot, ionized branch corresponding to large accretion
rates (outburst phase) and (2) a cold, neutral branch corresponding to
low accretion rates (quiescent phase).

Global time-dependent calculations of the thermal limit cycle show
that, in order to reproduce observed DN outbursts, viscosity on the
cold, neutral, branch must be much smaller than on the hot, ionized,
branch ($\alpha_{\rm cold} \approx 0.01 << \alpha_{\rm hot} \approx
0.1$; e.g., Smak 1984).  For $\alpha_{\rm cold} \approx \alpha_{\rm
hot}$, the disk only undergoes small amplitude luminosity
fluctuations, rather than the large amplitude outbursts observed.
Gammie \& Menou (1998; hereafter GM) showed that the magnetic Reynolds
number on the cold branch in DN disks is $Re_M \approx 3 \times 10^3$.
Numerical simulations of the magneto-rotational instability (Balbus \&
Hawley 1991; 1998) show that, at such low magnetic Reynolds numbers,
MHD turbulence and its associated angular momentum transport are
significantly reduced (Hawley, Gammie, \& Balbus 1996; Fleming, Stone,
\& Hawley 2000).  Menou (2000) used Fleming et al's calculation of
angular momentum transport as a function of $Re_M$ in a time-dependent
numerical model for DN outbursts.  He found that, as proposed by GM,
MHD turbulence dies away in quiescence.  This {interplay between
the thermal ionization instability and the magneto-rotational
instability} provides a physical motivation for the observational
inference that $\alpha_{\rm cold} \ll \alpha_{\rm hot}$; it suggests
that either accretion completely stops in quiescence or an additional,
{less efficient}, angular momentum transport mechanism operates
(perhaps spiral density waves excited by the companion; Spruit 1987;
Menou 2000).

Thin accretion disks around supermassive black holes in active
galactic nuclei are also expected to have regions with $T_c \sim 10^4$
K, and are thus also subject to the thermal ionization instability
(e.g., Lin \& Shields 1986; Cannizzo 1992; Siemiginowska, Czerny, \&
Kostyunin 1996).  Applications of the disk instability model to AGN
have, however, received less attention than those to DN and SXTs; this
is no doubt because of the much longer recurrence times, which are
observationally inaccessible in individual sources.  The thermal
ionization instability {could}, however, significantly influence
the luminosity function of AGN, perhaps accounting for the short
``duty cycle'' of bright Seyferts and quasars (e.g., Mineshige \&
Shields 1990; Siemiginowska \& Elvis 1997).  In addition, it could
lead to a radially varying accretion rate in the disk, complicating
comparisons between theoretical models and observed broad band
spectra.  Understanding the role of the ionization instability and
other sources of large amplitude variability in AGN is thus of
considerable importance.

In this {paper} we generalize GM's analysis and argue that, in
contrast to SXTs and DN, the gas {\it is well coupled} to the magnetic
field in AGN disks even on the cold branch of the thermal limit cycle
(quiescence).  This is essentially because the disks in AGN are
physically much larger than those in DN; for a fixed resistivity, the
magnetic Reynolds number is thus much larger.  Our analysis implies
that the thermal ionization instability is unlikely to be a
significant source of large amplitude variability in AGN.

{Accretion disks around supermassive black holes also differ from
disks in DN and SXTs in that they may develop both local and global
gravitational instabilities, which can lead to efficient angular
momentum transport (e.g. Shlosman, Begelman \& Frank 1990; Kumar 1999;
Gammie 2000).  In this paper, we argue that the degree to which 
the magneto-rotationally and gravitationally unstable portions
of the accretion disk overlap is important for understanding time-dependent
accretion in AGN.}

We proceed as follows.  We first calculate the structure of AGN disks
in regions subject to the thermal ionization instability (\S2).  We
then estimate the importance of resistivity, ambipolar diffusion, and
Hall currents in these regions (\S3).  In \S4 we reassess the role of
the ionization instability in AGN.  {In \S5 we discuss the
interplay between MHD (un)stable\footnote{Throughout this {paper},
we use ``MHD unstable'' to mean unstable to developing MHD turbulence
via the magneto-rotational instability.} and gravitationally
(un)stable regions of AGN accretion disks, and in \S 6 we conclude.}

\section{The Local {Thermal} Structure of AGN Disks}

We consider a gaseous, thin accretion disk around a supermassive black
hole (BH) of mass $M$.  We ignore X-ray irradiation in calculating
thermal equilibria (but see \S5).\footnote{X-ray irradiation can lead
to higher temperatures than considered below; this can substantially
change the thermal ionization instability (e.g., Dubus et al. 1999).}
The local thermal equilibria of the disk are then found by equating
the local viscous dissipation rate $Q^+$ to the radiative cooling rate
$Q^-$, where $Q^-$ is a function of the disk opacity. Such equilibria
have previously been calculated by several authors for AGN disks (see,
e.g., Mineshige \& Shields 1990; Cannizzo 1992; Siemiginowska et
al. 1996). As for disks in DN and SXTs, these equilibria are thermally
and viscously unstable in regions of partial ionization, corresponding
to $T_c \sim 10^4$~K.

\begin{figure}
\plotone{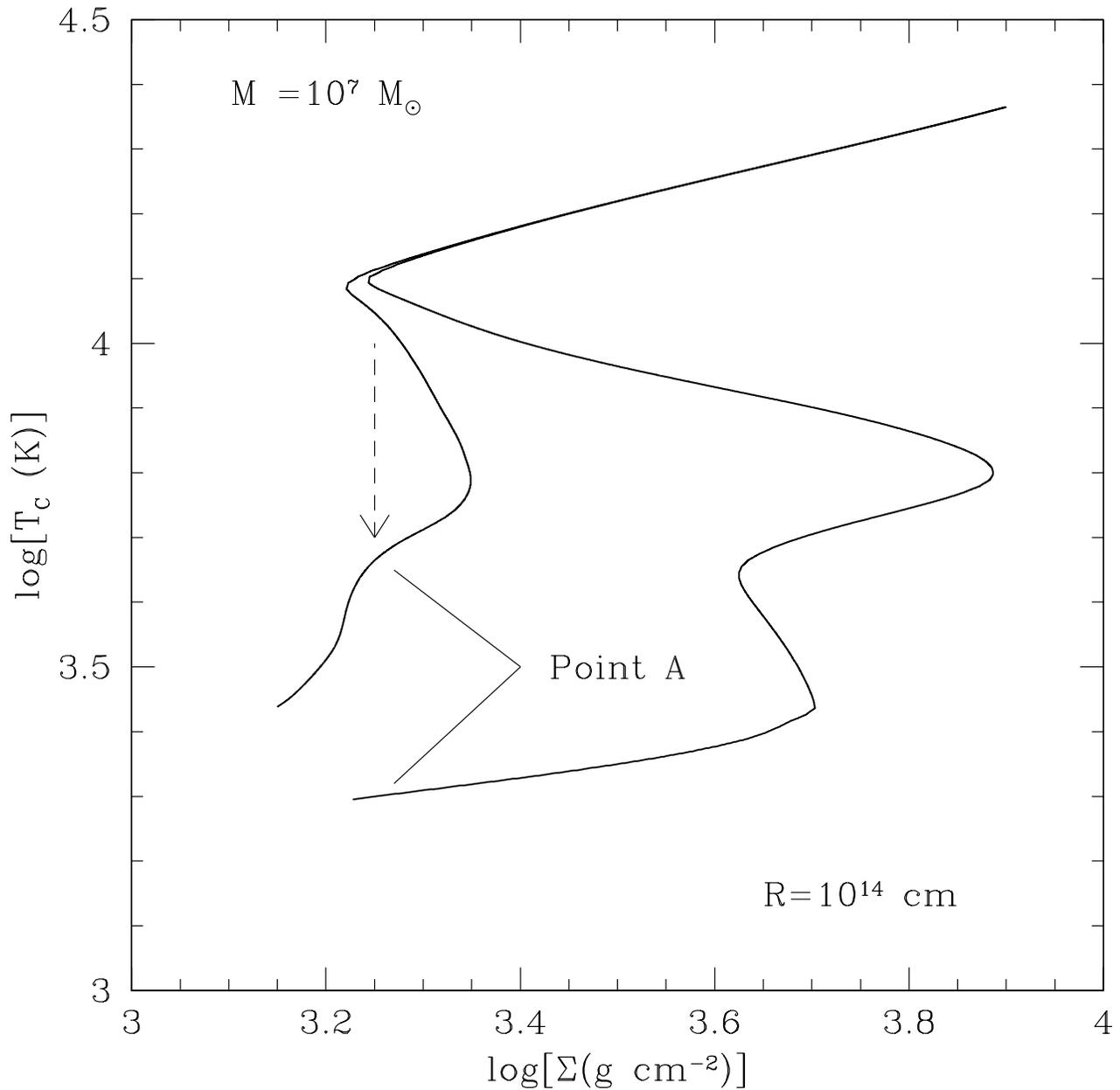}
\caption{Example ``S-curves'' (curves of thermal equilibria in a
surface density - central temperature ($\Sigma$-$T_c$) diagram) for a
thin accretion disk around a supermassive black hole. The disk annulus
is located at $R=10^{14}$~cm around a black hole of mass $M = 10^7
M_\odot$. For the upper S-curve, the values of the viscosity parameter
are $\alpha_{\rm hot}=\alpha_{\rm cold}=0.1$, while for the lower
S-curve $\alpha_{\rm hot}=0.1 \gg \alpha_{\rm cold}=0.01$. The point
labeled A on each of the lower branches corresponds to the coldest
state reached by the annulus during its thermal limit cycle evolution;
this is where the gas is {\it least} coupled to the magnetic field. 
\label{fig:scurve}}
\end{figure}

Figure~\ref{fig:scurve} shows two thermal equilibrium curves
(``S-curves'') for an annulus located at $R=10^{14}$~cm around a BH of
mass $M = 10^7 M_\odot$; for the upper S-curve, $\alpha_{\rm hot} =
\alpha_{\rm cold} = 0.1$, while for the lower S-curve, $\alpha_{\rm
hot} = 0.1 \gg \alpha_{\rm cold} = 0.01$.  The thermal equilibria have
been found by calculating a grid of models for the disk vertical
structure, for different values of the surface density $\Sigma$ and
the central temperature $T_c$, at the radius of interest (Hameury et
al. 1998).  Equilibria correspond to ($\Sigma$,$T_c$) pairs such that
$Q^+ = Q^-$.

The middle, inverted sections of the S-curves are where the gas in the
disk is partially ionized and unstable.  Any disk annulus ostensibly
in this state instead undergoes a limit cycle in which it
time-dependently evolves between the two stable (hotter and colder)
thermal equilibria.

Point A in Figure 1 represents the lowest temperature reached by any
unstable annulus during the thermal limit cycle.  This point is of
interest because it is the place in the limit cycle where the gas in
the disk is {\it least} coupled to magnetic fields (it has the
smallest ionization fraction in LTE).  As a conservative estimate, we
use $T_c \approx 2000$ K for the central temperature of the unstable
annulus at point A.  This is appropriate for $\alpha_{\rm cold} =
0.01$.\footnote{This result can be checked by solving the
vertically-averaged equation $Q^+ = (9/8) \nu \Sigma \Omega^2 = Q^-=4
\sigma T_c^4/(3 \Sigma \kappa)$ for an optically-thick disk, where
$\nu$ is the kinematic viscosity and $\kappa$ is the opacity (which
can be obtained from a standard opacity table for the range of density
and temperature of interest).  It is also interesting to note that the
temperature $T_A \simeq 2000$~K at point A on the S-curve is, for
$\alpha_{\rm cold} \approx 0.01$, a generic prediction for any black
hole mass and at any radius in the disk (see, e.g., GM for an
identical result in the case of a DN disk).}  If, as we argue in
\S3-4, MHD turbulence is responsible for accretion in AGN disks in the
vicinity of point A, $\alpha_{\rm cold} \approx \alpha_{\rm hot}
\approx 0.1$ would be more appropriate (see Hawley 2000 for global MHD
simulations which give $\alpha \approx 0.1$). This only strengthens
our conclusions.

\section{Coupling of the Gas to the Magnetic Field}

GM showed that $Re_M \approx 3 \times 10^3$ at point A in DN disks;
here we generalize their analysis and assess the coupling of gas to
magnetic fields in unstable annuli of accretion disks around compact
objects of arbitrary mass.

The importance of ohmic diffusion can be estimated using the magnetic
Reynolds number: \beq Re_M = {c_s H \over \eta}, \label{rem} \eeq
where $H \approx c_s/\Omega$ is the scale height of the disk, $c_s$ is
the sound speed $\eta = c^2 m_e \nu_{en}/(4 \pi n_e e^2)$ is the
resistivity, $\nu_{en} \approx 8.3 \times 10^{-10} T^{1/2} n_n$
s$^{-1}$ is the electron-neutral collision frequency (Draine, Roberge,
\& Dalgarno 1983), and $n_n (n_e)$ is the number density of neutrals
(electrons).  For the relatively low density disks of interest in AGN,
ambipolar diffusion can be more important than resistive diffusion.
Its importance can be estimated using the dimensionless number (e.g.,
Blaes \& Balbus 1994) \beq Re_A = {\nu_{ni} \over \Omega}.
\label{rea} \eeq $Re_A$ measures the ratio of the neutral-ion collision 
frequency, $\nu_{ni}$, to the orbital frequency in the disk, $\Omega$;
for $Re_A \gg 1$, the neutrals are well coupled to the magnetic field
by collisions with the ions.  In equation (\ref{rea}), $\nu_{ni} =
\gamma \rho_i$, $\gamma \approx 3.5 \times 10^{13}$ cm$^3$ s$^{-1}$
g$^{-1}$ is the drag coefficient for ion-neutral
collisions\footnote{We take $m_i = 30 m_H$ and $m_n = 2.33 m_H$, where
$m_i$ is the ion mass, $m_n$ is the neutral mass, and $m_H$ is the
mass of the hydrogen atom.}  (Draine et al.  1983), and $\rho_i$ is
the ion mass density.

To estimate $Re_M$ and $Re_A$, we use the physical conditions in the
limit cycle least favorable to MHD turbulence, namely those at point
A; in particular, we use $T_A \approx 2000$ K, a somewhat conservative
value (\S2).  From our numerical calculations of S-curves, we find
that the surface density at point A can be approximately fit by (see
also Cannizzo 1992; Siemiginowska et al. 1996) \beq \Sigma_A \approx
\Sigma_0 \ \alpha_{0.1}^{-3/4} \ r \ m_9^{2/3}, \label{sig} \eeq where
$\Sigma_0 \approx 1000$ g cm$^{-2}$, $\alpha_{0.1} = (\alpha/0.1)$,
$r$ is the radius of the unstable annulus in Schwarzschild units, and
$m_9 = (M/10^9 M_\odot)$.  Finally, for a given rate at which mass is
fed to the accretion flow at large radii ($\dot M$, in $\mpy$), the
radius around which the ionization instability sets in is given by \beq
r_A \approx r_0 \ \dot M^{0.4} \ m_9^{-2/3},
\label{ra} \eeq where $r_0 \approx 300$.  Equations (\ref{sig}) and
(\ref{ra}) can be roughly reproduced analytically by fixing $T \sim
10^4$ K in a Shakura-Sunyaev disk (thus isolating the portions of the
disk where hydrogen starts being partially ionized).  It is also
interesting to note that equations (\ref{sig}) and (\ref{ra}) are
applicable to DN and SXTs as well as AGN, with $m_9$ then $\sim
10^{-9}$, $\dot M \sim 10^{-9} \mpy$, and $r_A \sim 10^5$.

Using the Saha equation for a solar metallicity gas, we find that, in
LTE, the ionization fraction at $T_A \approx 2000$~K in the thermally
unstable annulus is $x_e \equiv n_e/n_n \approx 10^{-6}-10^{-7}$ for
$\dot M \ \epsilon \ [1,10^{-10}] \mpy$.\footnote{The neutral density
at point A in the unstable annulus is $n_n \sim 3 \times 10^{15} \dot
M^{-0.2}$ cm$^{-3}$.  From the Saha equation, $x_e \propto n_n^{-1/2}
\propto \dot M^{0.1}$.  Thus, at fixed $T_c$, the $\approx 10$ order
of magnitude change in $\dot M$ from DN to luminous AGN implies a
change in $x_e$ by a factor of $\approx 10$. Larger values of $x_e$
are expected in high metallicity AGN disks because $n_e$ is dominated
by contributions from metals with small ionization potentials.}
Substituting equations (\ref{sig}) and (\ref{ra}) into equations
(\ref{rem}) and (\ref{rea}) we then find the {\it smallest} values of
$Re_M$ and $Re_A$ in the thermal limit cycle (obtained at point A):
\beq Re_M \approx 10^8 \left({x_e \over 10^{-7}}\right) \left({r_0
\over 300}\right)^{3/2} \dot M^{0.6}
\label{remfinal} \eeq and \beq Re_A \approx 3 \times 10^7 \alpha^{-3/4}_{0.1} 
\left({x_e \over 10^{-7}}\right) \left({r_0 \over 300}\right)
\left({\Sigma_0 \over 1000}\right) \dot M^{0.4}. \label{reafinal} \eeq
For a given accretion rate, equations (\ref{remfinal}) and
(\ref{reafinal}) determine the coupling of the gas to the magnetic
field in the annulus which is unstable to the ionization instability;
this annulus lies at a radius given by equation (\ref{ra}).  Note that
$Re_M$ and $Re_A$ are independent of black hole mass at fixed $\dot
M$.

Hall currents can significantly modify the dynamics of low density,
weakly-ionized disks (e.g., Wardle 1999; Balbus \& Terquem 2000).
Their importance can be measured using the ion Hall parameter,
$\beta_i \equiv \omega_i/\gamma \rho_n$, where $\omega_i$ is the ion
cyclotron frequency.  For $\beta_i \gg 1$ the ambipolar diffusion
limit applies, while for $\beta_i \ll 1$ Hall currents are important.
Assuming equipartition magnetic fields, we estimate $\beta_i \approx
0.3 \dot M^{0.1} (\Sigma_0/1000)^{-1/2}$ in the unstable annuli of
interest.  The role of Hall currents in modifying MHD turbulence is
determined by $Re_H \approx \beta_i Re_A$; since $\beta_i \sim 1$,
$Re_A$ is a reasonable proxy for $Re_H$: if $Re_A \gg 1$, then $Re_H
\gg 1$, {MHD turbulence is fully developed}, and Hall currents are
unimportant.  By contrast, if $Re_A \sim 1$ MHD turbulence is
modified (and Hall currents must be considered {because of their
potentially destabilizing effects}).

\section{Implications}

MHD turbulence and its associated angular momentum transport are
expected to be suppressed for $Re_M$ and/or $Re_A$ less than critical
values $\equiv Re_M^c$ and $Re_A^c$.  Numerical simulations with
explicit resistivity are consistent with $Re_M^c \approx 10^4$ (Hawley
et al. 1996; Fleming et al. 2000), while simulations of ion-neutral
disks give $Re^c_A \approx 100$ (Hawley \& Stone 1998; see also Mac
Low et al. 1995; {Brandenburg et al. 1995}).

Equations (\ref{remfinal}) and (\ref{reafinal}) then show that there
is an important difference between accretion in DN and SXTs, where
$\dot M \sim 10^{-9} \mpy$, and accretion in luminous AGN, where $\dot
M \sim 0.01-10 \mpy$.  As shown by GM, $Re_M \lsim Re_M^c$ in DN and
SXTs.  MHD turbulence is thus suppressed on the ``cold'' branch of the
thermal limit cycle.

By contrast, for reasonable $\dot M$ in AGN, $Re_M \gg Re_M^c$ and
$Re_A \gg Re_A^c$. This implies that the disk is MHD-turbulent
throughout the thermal limit cycle.  In contrast to DN and SXTs, the
efficiency of angular momentum transport should then be comparable on
the hot and cold branches ($\alpha_{\rm cold} \approx \alpha_{\rm
hot}$).  In this case global time-dependent calculations show that the
thermal ionization instability is probably not a significant source of
time-dependent accretion in AGN; it leads to luminosity
``flickering,'' rather than large amplitude outbursts analogous to DN
and SXTs (e.g., Smak 1984; Mineshige \& Shields 1990).

The analysis presented {above} does not, of course, preclude that
accretion in AGN is highly time dependent; it does, however, suggest
that the thermal ionization instability, as applied to DN and SXTs, is
not relevant to AGN.  Radiation-pressure induced thermal and viscous
instabilities at small radii may be an important source of time
dependence in AGN accretion disks (Lightman \& Eardley 1974; Piran
1978).  Another, less well explored, possibility lies in the interplay
between MHD (un)stable and gravitationally (un)stable portions of the
accretion disk.  We briefly address several of the relevant issues
{below}; much more theoretical work on these problems is needed.

{

\section{Magneto-rotationally and Gravitationally Unstable Regions}

\subsection{Magneto-rotational Stability}
}

An obvious shortcoming of the argument in \S3 is that it is local.  We
examined annuli in thin accretion disks which are subject to the
thermal ionization instability and found that they are always MHD
turbulent so long as $\dot M \gg 10^{-7} \mpy$.  At radii somewhat
larger than the thermally unstable region, however, the disk is
everywhere stable, but on the {\it cold} branch of the S-curve in
Figure 1; if viscous dissipation is the only source of heating, the
central temperature in the disk eventually drops to $\ll 1000$ K.  In
LTE the ionization fraction then becomes very small and MHD turbulence
dies away.

X-rays and cosmic-rays may provide sufficient heating and/or
nonthermal ionization to maintain magnetic coupling in the disk at
large radii.  Blaes \& Balbus (1994) estimated $Re_M \gg 1$ and $Re_A
\sim 1$ from cosmic-ray ionization in the parsec scale circumnuclear
disk in our Galaxy.  Simple estimates of X-ray ionization yield
similar results.

As a concrete example, consider the $\approx 0.1-0.25$ pc masing disk
around the $M = 3.6 \times 10^7 M_\odot$ black hole in NGC 4258
(Miyoshi et al. 1995).  Neufeld \& Maloney (1995; NM) proposed that
the masing region is strongly irradiated by the central X-ray source,
due to a significant warp in the disk.  For a disk illuminated at an
angle cos$^{-1} \mu$ to the normal direction, the incident {X-ray}
flux on the disk is $f \approx (\mu L_X)/(4 \pi R^2)$, where $L_X
\approx 4 \times 10^{40}$ ergs s$^{-1}$ for NGC 4258 (Makishima et
al. 1994) and $\mu$ may be as large as $\approx 0.2$ (NM; Herrnstein
1997).  The volume ionization rate {in the bulk of the disk} due
to the incident X-ray flux is $\xi \approx {\rm min}(1,\tau) f /(H
E_i)$ cm$^{-3}$ s$^{-1}$, where $E_i \approx 37$ eV {is the
typical energy required to generate one secondary electron}
(Glassgold, Najita, \& Igea 1997), $\tau = \mu^{-1} \Sigma/\Sigma_c$
is the shielding optical depth of the disk, and $\Sigma_c \approx 1$ g
cm$^{-2}$ is the column density corresponding to order unity optical
depth for hard X-rays $\gsim 10$ keV ({the factor ${\rm
min}(1,\tau)$ is included to take into account the possibility of a
disk which is optically-thin to X-rays}).  NM find $n_n \approx 3
\times 10^7$ cm$^{-3}$ in the masing region and so $\Sigma \approx
0.1$ g cm$^{-2}$.  Balancing X-ray ionization against dissociative
recombination at a rate $8.7 \times 10^{-6} n^2_e T^{-1/2}$ cm$^{-3}$
s$^{-1}$ yields $n_e \approx 300$ cm$^{-3}$ and thus $x_e \approx
10^{-5}$.  From equations (\ref{rem}) and (\ref{rea}), this
corresponds to $Re_M \sim 10^{11}$ and $Re_A \sim 10^3$.  Somewhat
counterintuitively, the disk in NGC 4258 may be both masing and
MHD-turbulent!  Observational support for this conclusion is provided
by the work of Wallin, Watson, \& Wyld (1998, 1999); they showed that
the {maser} spectral line features observed in NGC 4258 can be
reproduced reasonably well by including a turbulent velocity field in
a thin Keplerian disk model.

In order to account for the absence of masing at radii $\lsim 0.1$ pc,
NM propose that X-ray irradiation ``shuts off'' inside $\approx 0.1$
pc (because the warp decreases with radius, $\mu \rightarrow 0$).  If
correct, MHD turbulence probably ceases to be a viable angular
momentum transport mechanism (cosmic-ray ionization alone yields $Re_M
\sim 10^7 \xi_{17}^{1/2}$, but $Re_A \sim 0.1 \xi_{17}^{1/2}$, where
$\xi = 10^{-17} \xi_{17}$ s$^{-1}$ is the cosmic-ray interaction rate
per hydrogen atom).\footnote{The ion Hall parameter is $\beta_i
\approx 10^3$ {assuming equipartition fields}, so Hall currents
should not change this result.  Scattered X-rays from the vicinity of
the AGN could, however, provide significant ionizing photons; their
importance is very uncertain.}

This illustrates a more general worry; the role of nonthermal heating
and ionization is quite uncertain, depending on, e.g., the geometry of
the disk (warps, flaring, etc.), the presence of X-ray scatterers, and
the possibility of an anomalously large cosmic-ray flux due to a
nearby jet.  It is possible that many parsec scale disks in AGN are
{\it not} magnetically well coupled.  In this case gravitational
instability is the most promising angular momentum transport
mechanism.

{
\subsection{Gravitational Stability}
}

Thin disks are {\it locally} gravitationally unstable for $Q \lsim 1$,
where $Q = c_s \Omega/(\pi G \Sigma)$.  The nonlinear outcome of local
gravitational instability depends on the ratio of the cooling time to
the rotational period of the disk (Shlosman \& Begelman 1989; Shlosman
et al. 1990; Gammie 2000).  For long cooling times, the disk reaches a
steady state with efficient angular momentum transport produced by
``gravito-turbulence'' (Shlosman et al. 1990; Gammie 2000).  For
parsec-scale disks in AGN, however, the cooling time is generally much
shorter than the rotational period of the disk; in this case the disk
fragments, and probably undergoes efficient star formation (Shlosman
\& Begelman 1989; Gammie 2000).  {It is unlikely that local
gravitational instability produces significant accretion in this
regime.}

For $M_{\rm disk} \approx 2 \pi \Sigma R^2 \gsim M$, however, the disk
is {\it globally} gravitationally unstable.  Using $\dot M = 2 \pi
\Sigma R |v_r|$ and $|v_r| \approx \alpha c^2_s/v_\phi$, where
$v_\phi$ is the (Keplerian) rotational velocity, this condition can be
rewritten as (Shlosman \& Begelman 1989) \beq \dot M \gsim {\alpha
c^2_s v_\phi \over G} \approx {10^{-2} \alpha_{0.1} T_{3} v_{100}}
\mpy,
\label{global} \eeq where $T = 10^3 T_{3}$ K and $v_{\phi} = 100 v_{100}$ 
km s$^{-1}$.

If nonthermal heating and ionization are insufficient to maintain MHD
coupling, disks will become MHD stable if $T_3 \lsim 1$.  For luminous
AGN with $\dot M \gsim 10^{-2} \mpy$, however, the disk is globally
gravitationally unstable in this region (eq. [\ref{global}]).  It is
therefore likely that bar formation in the gaseous disk leads to
significant angular momentum transport -- and thus accretion -- on a
dynamical timescale (Shlosman et al. 1990).  At smaller radii, where
$T_3 \gsim 1$, the disk is both MHD and gravitationally unstable; the
dynamics in this regime has been poorly explored but may be accessible
with numerical simulations.  More generally, in luminous AGN the
accretion disk probably evolves from gravitationally unstable and MHD
stable at large radii to gravitationally unstable, MHD unstable, and
finally to gravitationally stable, MHD unstable at the smallest radii.
{It is rather uncertain whether these transitions occur at
constant $\dot M$.}

In low-luminosity AGN with $\dot M \lsim 10^{-2} \mpy$ equation
(\ref{global}) shows that the disk is globally gravitationally stable
where it is MHD stable; for $\dot M \lsim 10^{-3} \mpy$ the disk has
$Q \gsim 1$ and is locally stable as well; this raises the interesting
possibility of an MHD and gravitationally stable ``dead zone'' in
which accretion may proceed very inefficiently (or perhaps not at
all).\footnote{Accretion may proceed in an ionized surface layer
analogous to Gammie's (1996) proposal for T-Tauri disks.}  Mass will
build up in this region until $M_{\rm disk} \gsim M$, until the disk
warp evolves so as to provide significant X-ray ionization, or until a
global hydrodynamic instability sets in (analogous to GM's proposal
for DN).  It is possible that such piling up of mass is presently
occurring in many nearby galaxies.  In particular, returning again to
NGC 4258, NM's masing model gives $\dot M \approx 10^{-5} \alpha_{0.1}
\mpy$ and thus a very gravitationally stable disk: $M_{\rm disk} \sim
100 M_\odot \ll M$ and $Q \approx 100 \gg 1$ (see Gammie, Narayan, \&
Blandford 1999 for alternate $\dot M$ estimates).  If X-ray
irradiation and MHD turbulence indeed shut off for $R \lsim 0.1$ pc,
it is unlikely that significant accretion occurs at these radii.  It
is then also unlikely that NM's accretion rate estimate is relevant
for the energy-producing portions of the accretion flow at small
radii.

\

{
\section{Conclusion}

We have studied the combined role of the thermal ionization,
magneto-rotational and gravitational instabilities in thin accretion
disks around supermassive black holes.

In contrast to dwarf novae and soft X-ray transients, we find that
there is no physical motivation for a reduced efficiency of angular
momentum transport when AGN disks become neutral via the thermal
ionization instability. This is because even the predominantly neutral
gas is well coupled to the magnetic field ($Re_M \gg 1$ and $Re_A \gg
1$).  MHD turbulence due to the magneto-rotational instability should
thus be present and similar to that in an ionized plasma. In this
case, numerical simulations have shown that the thermal ionization
instability leads to small amplitude luminosity ``flickering,'' rather
than large amplitude outbursts analogous to DN and SXTs.

We propose, however, that accretion in AGN can be time-dependent
because of the interplay between the magneto-rotational and
gravitational instabilities, both of which are viable candidates for
transporting angular momentum in AGN accretion disks. The hot inner
regions of the disk are preferentially subject to the
magneto-rotational instability, while the cold outer regions are prone
to developing gravitational instabilities.  In high luminosity systems
(accretion rate $\dot M \gsim 10^{-2} \mpy$), the magneto-rotationally
and gravitationally unstable regions of the accretion disk overlap,
while in lower luminosity systems ($\dot M \lsim 10^{-3} \mpy$) they
may not. In systems where these regions do not overlap, such as nearby
low-luminosity galactic nuclei on scales of $\sim 0.1-1$ parsec, there
is no well understood process for removing angular momentum from
accreting gas. Mass may therefore be temporarily building up in a
non-accreting ``dead-zone.''

}

\section*{Acknowledgments}

We thank Bruce Draine, Pawan Kumar, Sergei Nayakshin, {the referee, }
and especially Charles Gammie for useful comments or
discussions. Charles Gammie provided a Mathematica routine to
calculate ionization fractions.  Support for this work was provided by
NASA through Chandra Fellowship grant PF9-10008 (to EQ) and PF9-10006
(to KM) awarded by the Smithsonian Astrophysical Observatory for NASA
under contract NAS8-39073.

\clearpage

\end{document}